\documentstyle[floats,aps,epsf,preprint]{revtex}
\tightenlines
\begin{document}
\draft
\preprint{BI-TP-97/12, SU-4240-661}
 
\title{Numerical Observation of a Tubular Phase in Anisotropic Membranes}

\author{Mark Bowick,$^{1}$ Marco Falcioni$^{1}$ and Gudmar Thorleifsson$^{2}$}
 
\address{$^1$Physics Department, Syracuse University, 
 Syracuse NY 13244-1130, USA\\}
 
\address{$^2$Fakult\"{a}t f\"{u}r Physik, Universit\"{a}t Bielefeld,
D-33615, Bielefeld, Germany\\} \maketitle
\begin{abstract}
We provide the first numerical evidence for the existence of a
tubular phase, predicted by Radzihovsky and Toner (RT), for
anisotropic tethered membranes without self-avoidance.  Incorporating
anisotropy into the bending rigidity of a simple model of a tethered
membrane with free boundary conditions, we show that the model indeed
has two phase transitions corresponding to the flat-to-tubular and
tubular-to-crumpled transitions. For the tubular phase we measure the
Flory exponent $\nu_F$ and the roughness exponent $\zeta$.  We find
$\nu_F=0.305(14)$ and $\zeta=0.895(60)$, which are in reasonable
agreement with the theoretical predictions of RT --- $\nu_F=1/4$ and
$\zeta=1$.
\end{abstract}
\normalsize 

Tethered membranes are 2-dimensional regularly triangulated surfaces
fluctuating in 3 dimensions.  Their behavior at thermal equilibrium is
governed by the elastic and bending moduli, which correspond to the
in-plane and out-of-plane rigidities respectively. Isotropic tethered
membranes have been studied extensively, and it is known that they
have a crumpled high temperature phase and a flat low temperature
phase separated by a continuous phase transition, the {\em crumpling}
transition \cite{revs,lh94,KN}.

In a recent paper Radzihovsky and Toner (RT) have shown that
anisotropy can radically change the nature of the phase diagram for
tethered membranes \cite{bib1,bib2}.  In particular a remarkable and
completely unanticipated new phase of non-self-avoiding ({\em
phantom}) tethered membranes {---} the {\em tubular} phase {---} is
predicted.  The tubular phase is characterized by the presence of
long-range orientational order in one direction only {---} in the
transverse directions it is crumpled.  Furthermore it is expected that
any degree of anisotropy will eliminate the direct crumpling
transition from the flat to crumpled regimes and replace it by two
transitions {---} a low temperature flat-to-tubular transition and a
higher temperature tubular-to-crumpled transition.  Since it is very
likely that real anisotropic tethered membranes can be observed
experimentally, the rich structure of the phase diagram of these
systems is exciting.  There are several experimental realizations of
anisotropic membranes one could imagine.  Polymerized membranes with
in-plane tilt order are good candidates \cite{bib1}. It may also be
feasible to cross-link in an applied electric field DNA molecules
trapped in a fluid membrane \cite{bib1}.  Fluid membranes themselves
also exhibit anisotropic ``ripple'' phases \cite{bib3}.

A tethered membrane in the tubular phase is quite different from real
tubules, which are well defined cylindrical structures, such as lipid
tubules, protein microtubules and carbon nanotubes
\cite{tubul,tubules1,tubules2}. Fluid membranes with chiral order
exhibit tubular shapes \cite{chiral}.  In certain experimental
conditions these tubules display a wealth of interesting behavior such
as the pearling instability \cite{exp,pearls}.

The existence of an almost one-dimensional ordered phase is quite
remarkable and, at least for phantom membranes, rather delicate.  A
tubular phase is possible because of the inevitable transverse
stretching energy cost of bending fluctuations in the extended
direction.  Its existence is not trivial, though, because for physical
tethered membranes, with internal dimension 2 and embedding dimension
3, the stability is marginal.  The fluctuations of the tubule away
from a straight linear shape (height fluctuations) along the extended
direction are maximal with a corresponding roughness exponent of 1
\cite{bib1}.  For this reason it is imperative to check the existence
of stable tubules with careful numerical simulations.

In this paper we establish the existence of tubules by large-scale
Monte Carlo simulations of a discrete model of tethered membranes with
the topology of a disk and free boundary conditions.  The isotropic
version of this model has been extensively studied by us recently
\cite{bib4} and so we are well-equipped to assess the effects of
anisotropy.

We have chosen to implement the anisotropy in the bending rigidity
rather than the elastic moduli of the model. Hence the membrane
responds isotropically to in-plane stresses, but anisotropically to
out-of-plane bending. This is conceptually clearer for us and, from
the arguments given in \cite{bib1}, is equally good. Thus we choose
the bending rigidity on links in the $x$-direction ($\kappa_1$) of a
triangular lattice to be larger than the bending rigidity along links
in the other two directions ($\kappa_2$).  We study a system with
aspect ratio one ($L_x=L_y$), so that any anisotropy that develops is
inherent to the system rather than introduced {\em ab initio}. Different
critical exponents arise \cite{bib1} if one takes thermodynamic limits
with other aspect ratios (i.e. if one tunes $L_x$ to diverge as some
power of $L_y$).
\begin{figure}[t]
\epsfxsize= 5in \centerline{\epsfbox{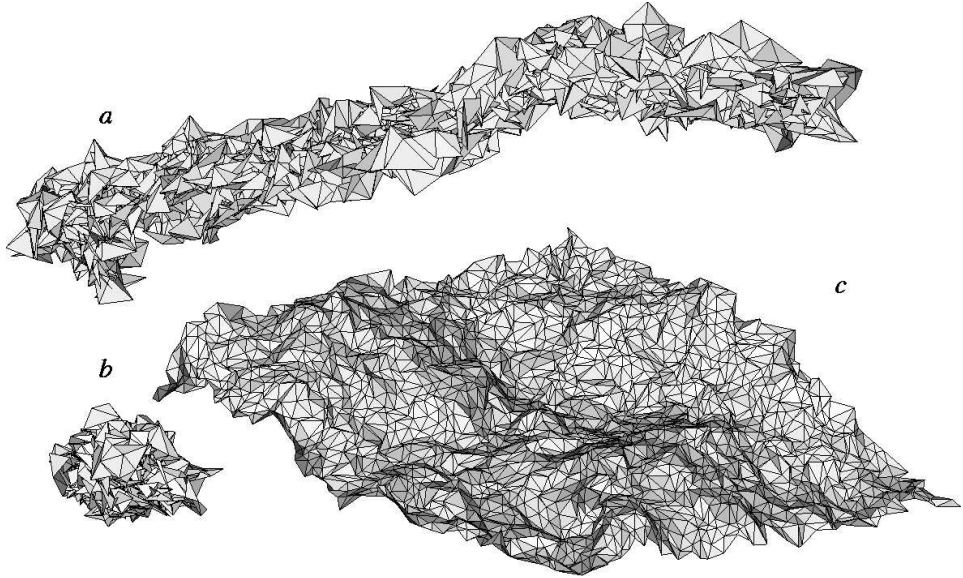}}
\caption{a) An example of a tubular membrane. This is for $L = 65$ and
$(\kappa_1,\kappa_2) = (2,0)$. b) An isotropic membrane ($L=46$) 
in the crumpled phase ($\kappa=0.5$) and c) in the flat phase 
($\kappa=1.1$).}
\label{fig1}
\end{figure}

By examining the specific heat along various paths in the $(\kappa_1,
\kappa_2)$ plane we find that the model indeed exhibits two distinct
transitions. The stronger transition is the one from the tubular to the
flat phase. The tubular phase is characterized by equilibrium
configurations which are extended in the $y$-direction and crumpled in
the transverse direction (see Fig.~\ref{fig1}a).  Given this evidence
for the existence of the tubular phase we proceed to measure the
appropriate critical exponents and to compare them with the
predictions of \cite{bib1}.

The exponents we focus on are the Flory exponent $\nu_F$ and the
roughness exponent $\zeta$.  The first gives the scaling of the
transverse radius of gyration $R^G_{\perp}$ (the tubule radius) with
system size $L$, $R^G_{\perp} \sim L^{\nu_F}$, and is predicted to be
$1/4$. The second relates the height fluctuations in the extended
($y$) direction to the system size $\langle h^2 \rangle = \langle
(h(x_{\perp},L) - h(x_{\perp},0))^2 \rangle \sim L^{2 \zeta}$.  The
theoretical predictions of RT for these exponents for square phantom
membranes are unambiguous.  They find $\nu_F=1/4$ and $\zeta=1$. The
maximal value for $\zeta$ corresponds to the previously mentioned
marginal stability of the tubular phase.

Let us now give the details of our simulation.  We analyze the
partition function
\begin{equation}
 Z \:=\: \int [\rm d{\bf r}] \: \delta ({\bf r}_{cm}) \:
 {\rm e}^{\textstyle - {\cal H}[\bf r] }
\label{eq11}
\end{equation}
where ${\rm \bf r} \in {\rm \bf R}^3$ are the embedding coordinates
and the delta function ensures that the center of mass motion is
eliminated. The discrete Hamiltonian ${\cal H}$ is composed of a
tethering potential, in the form of a Gaussian spring with vanishing
equilibrium spring length, and (two) bending energy terms in the form
of ``ferromagnetic'' interactions between nearest-neighbor normals
${\rm \bf n}_a$ to the surface faces:
\begin{eqnarray}
 {\cal H}[{\rm \bf r}] & = & \sum_{\langle\sigma \sigma^\prime\rangle} \left|
 {\rm \bf r}_{\sigma} 
 - {\rm \bf r}_{\sigma^{\prime}} \right|^2 \\ \nonumber
 & - & \kappa_1 {\sum_{\langle ab\rangle}}^{(x)} {\bf n}_a \cdot {\bf n}_b
 - \kappa_2 {\sum_{\langle ab \rangle}}^{(y)} {\bf n}_a \cdot {\bf n}_b \,.  
 \label{eq12}
\end{eqnarray}
The second sum is over links in the $x$-direction and the third sum
over the other two link directions.  Note that no elastic constant are
explicitly introduced; they are dynamically generated.  We refer to
\cite{bib4} for an extensive treatment of the properties of this model
for the flat phase of tethered membranes.

The Hamiltonian above was simulated using Monte Carlo methods for
triangular lattices of up to $65^2$ nodes.  The global shape of the
lattice is chosen to be square with free boundary conditions.  The
field configurations \{{\bf r}\} were updated using a unigrid
algorithm, which substantially reduces the auto-correlation times
compared to a simple Metropolis algorithm.  We performed typically
$10-20\times10^6$ sweeps per lattice volume and coupling {---} this
resulted in about $5-10\times10^3$ independent measurements.  The
simulations required $\approx 50,000$ CPU hours on an IBM RS/6000
computer.
\begin{figure}[t]
\epsfxsize= 4in \centerline{\epsfbox{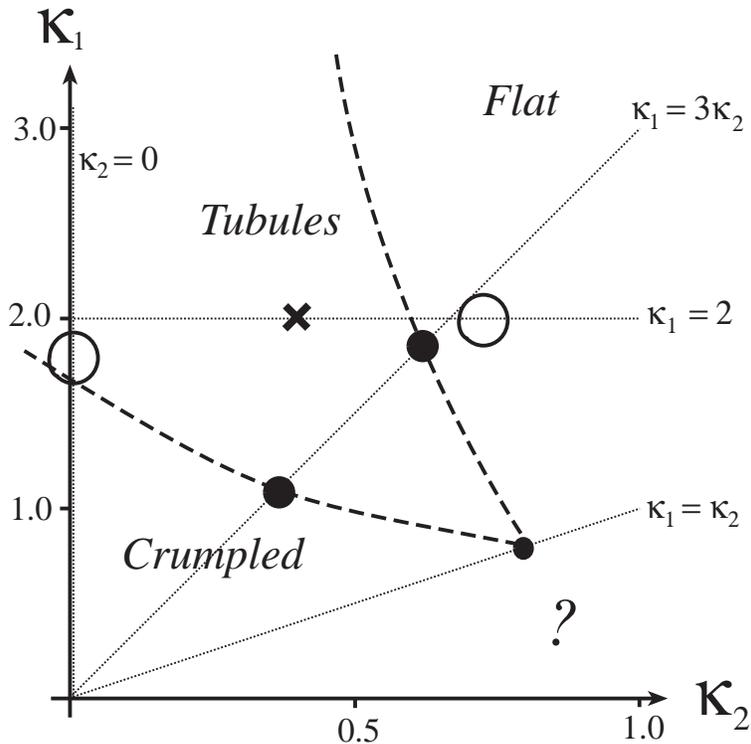}}
\caption{The phase diagram of an anisotropic tethered membrane.  The
 circles correspond to observed peaks in the specific heats $C_V^x$
 and $C_V^y$ (the filled ones are from larger lattices).  We performed
 simulations along the dotted lines with the cross indicating where we
 studied the tubular phase.}
\label{fig3}
\end{figure}

To explore the phase diagram predicted by RT we performed extensive
simulations along the line $(\kappa_1,\kappa_2) = (3\kappa, \kappa)$
(see Fig.~\ref{fig3}).  In addition we looked at the lines
$(\kappa,0)$ and $(2,\kappa)$ on smaller lattices.  To look for
evidence of a phase transition we measured the two bending energy
terms in the action $E_x$ and $E_y$, for a range of couplings
$\kappa$, and looked at their fluctuations or, more specifically, the
two respective specific heats:
\begin{equation}
C_V^i(\kappa) = \frac{\kappa^2}{L^2} \;
 \frac{\partial}{\partial \kappa} 
\langle E_i \rangle.
\end{equation}
\begin{figure}[t]
\epsfxsize= 4in \centerline{\epsfbox{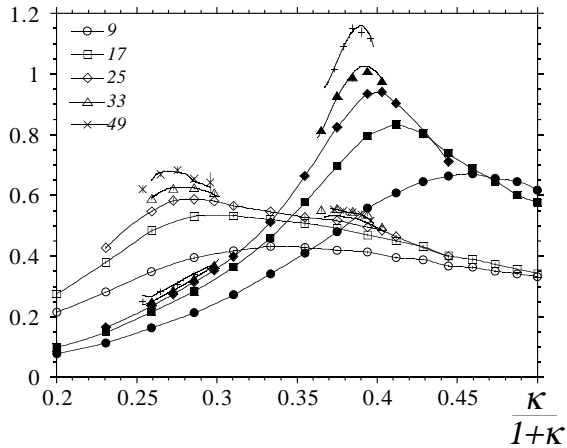}}
\caption{The measured values of the specific heats
 $C_V^x$ (open symbols) and $C_V^y$ (filled symbols).
 The interpolating lines are obtained using 
 multi-histogramming methods.}
\label{fig2}
\end{figure}

In Fig.~\ref{fig2} we show the measured values of $C_V^x$ and $C_V^y$
along the line $\kappa_1 = 3\kappa_2$.  This is for lattice sizes up
to $49^2$.  Both quantities show a divergent specific heat,
characteristic of a phase transition.  But what is remarkable is that
the peaks occur at {\it different} values of the bending rigidity,
signaling the existence of {\it two} distinct phase transitions.  As
the peak locations, which define the pseudo-critical couplings, shift
as the volume is increased, one may naturally ask if they merge in the
infinite volume limit.  To check that, we have fitted the
pseudo-critical couplings to the expected finite size behavior
\cite{bib5}: $\kappa^c_L = \kappa^c_{\infty} + c/L^{1/\nu}$.  For the
two peaks we get $\kappa^c_{\infty} = 0.36(2)$ and $\kappa^c_{\infty}
= 0.62(2)$, respectively, convincingly excluding the possibility of
them merging.  The corresponding values of the exponent $\nu$ are
$0.75(10)$ and $0.65(10)$, although we must caution that, for a
reliable estimate, bigger lattices are needed.  This in turn yields
the critical exponent $\alpha$ governing the divergence of the
specific heat.  Assuming the validity of hyper-scaling, $\alpha = 2 -
\nu d$, we find $\alpha \approx 0.5$ and 0.67, consistent with a
continuous phase transition.
    
One can also try to estimate $\alpha$ directly from the scaling of the
peak heights, although that is more difficult.  Preliminary estimates
yield a small positive number, again somewhat smaller for $C_V^y$.

Taken at face value, this might indicate that the tubular-to-crumpled
transition is somewhat weaker (slower divergence of the specific
heat), although simulations on larger lattices are needed to confirm
that the critical behavior of the two transitions really is different.
This work is in progress.  A detailed scaling theory of these two
transitions has been developed in \cite{bib2} and we plan to test this
theory once we have the very high statistics required for such a
comparison.

Performing a similar analysis along the other two lines in the
$(\kappa_1, \kappa_2)$ plane, although on smaller lattices, yields the
phase diagram shown in Fig.~\ref{fig3} \cite{bib6}.  This implies a
three phase structure, the usual high-temperature crumpled and low
temperature flat phases, together with the intermediate tubular phase
predicted in \cite{bib1}.

To investigate the nature of this phase we have performed extensive
simulations at the coupling $(\kappa_1=2,\kappa_2=0.4)$ on lattices
ranging up to $65^2$ in extent.  Direct evidence for the tubular
nature of the membrane is obtained by visual examination of
equilibrium configurations (Fig.~\ref{fig1}) and by measuring the
scaling of the three eigenvalues of the {\it shape tensor}
\cite{bib4}, defined as the off-diagonal part of the moment of inertia
tensor.  In a body-fixed frame these eigenvalues measure the extent of
the membrane along the associated principal axes.  We find that the
largest eigenvalue scales as $L^{2\nu_F}$, with $\nu_F \approx 0.988(11)$,
while the other two have much smaller $\nu_F$.  This indicates one
extended direction and one crumpled direction {---} i.e. a tubule.

A more detailed understanding of this tubular phase is obtained by
looking at the fluctuations of the zero-mode of the tubule height
$h_{rms}$, analogous to the height fluctuations of a flat membrane,
and the scaling of the width of the tubule $R^G_{\perp}$.
\begin{figure}[t]
\epsfxsize= 4in \centerline{\epsfbox{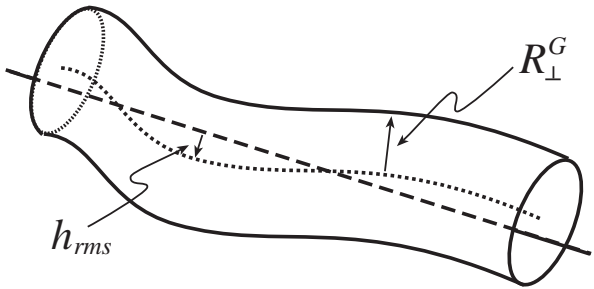}}
\caption{The definition of the fluctuations of the zero-mode 
$h_{rms}$ and the width of the tubule $R^G_{\perp}$.  
The dotted line indicates the one-dimensional shape of the tubule,
whereas the dashed one is the optimal straight line. } 
\label{fig4}
\end{figure}

The fluctuations of the zero mode are defined as the deviations of the
linear shape of the tubule from a straight line, whereas its width is
the average distance of points from this one-dimensional structure
(Fig.~\ref{fig4}).  We have measured this width in two different ways.  First,
we fit both a polynomial, of sufficiently high degree, and a straight
line to the membrane.  Usually a polynomial of degree $\approx 15$
produces stable results.  We then calculate the average distance of
the polynomial from the straight line and the average distance of all
points from the polynomial.  This was done for about 10,000
independent configurations of the membrane for each of the lattice
volumes $L=17, 25, 33, 49$ and $65$ (slightly fewer for the two
largest volumes).  The results were fit to the scaling predictions
\begin{equation}
h_{rms} \sim L^{\zeta} \;\;\; {\rm and}  
\;\;\;  {R^G_{\perp}} \sim L^{\nu_F}.
\end{equation}

Discarding the smallest lattice size we find $\zeta = 0.895(60)$ 
and $\nu_F = 0.305(14)$. These values compare quite  favorably
with the analytic continuum predictions of RT: $\zeta=1$ and
$\nu_F=1/4$.
Thus the combined analytic and numerical investigations
of tubules provide strong support for their existence and yield
consistent values for critical exponents.

The second method we use is to slice the tubule up as a salami,
defining a slice as the set of nodes having the same internal
$y$-coordinate.  Then the center of mass of the slice defines the
linear shape of the tubule. If we consider the shape tensor of the
ensemble of $\{x_{cm}\}$, the two larger eigenvalues scale as the
width of the tubule (squared).  We found that the two largest
eigenvalues scale with Flory exponents $\nu_F = 0.297(8)$ and
$0.241(12)$, respectively.  The smallest eigenvalue scales
logarithmically with the volume ($\lambda_{min} \sim \log(L)$), which
implies that the salami slices have vanishing thickness in the
infinite volume limit, thereby justifying the use of the method.  The
scaling of the average distance of the c.m.\ of the salami slices from
a straight line fit of the whole tubule yields the roughness exponent
$\zeta = 0.94(4)$, in reasonable agreement with the previously
obtained value.  It should be noted that the errors quoted are from
the quality of the fit; the actual uncertainty is dominated by finite
size effects.

In this paper we have established numerically for the first time the
existence of the remarkable tubular phase of anisotropic tethered
membranes predicted in \cite{bib1}. Furthermore we measure the Flory
exponent and the roughness exponent and find values not very far from
those expected analytically. This should stimulate further work on the
rich phase diagram of anisotropic tethered membranes.

There are two important directions to extend the simulations described
in this letter.  First one should investigate the anisotropic, rather
than square membrane, scaling limits studied in \cite{bib1,bib2}.  For
these systems the critical exponents are different.  Even more
exciting, perhaps, would be an investigation of self-avoiding
anisotropic membranes.  Self-avoidance causes extensive swelling of
the membrane in the transverse direction and also lowers the roughness
exponent $\zeta$ in the extended direction, as described by RT
\cite{bib1}.  Since self-avoidance is irrelevant in the extended
direction, in the tubular phase, it need only be implemented in the
transverse direction.  Such simulations should therefore be less
demanding than analogous simulations for isotropic membranes.  Precise
predictions exist for all relevant exponents for self-avoiding
anisotropic membranes in the tubular phase \cite{bib1,bib2,bib7}.

We would like to thank Leo Radzihovsky, Emmanuel Guitter and David
Nelson for helpful discussions.  We are grateful to NPAC (North-East
Parallel Architecture Center) and the Cornell Theory Center for use of
their computational facilities.  The research of MB and MF was
supported by the Department of Energy U.S.A.\ under contract
No.~DE-FG02-85ER40237 and that of MF by a Syracuse University Graduate
Fellowship.  The research of GT was supported by the Alexander von
Humboldt Stiftung and the Deutsche Forschungsgemeinschaft.


\begin{references}
\bibitem{bib5}
To analyze the peaks we have interpolated between the measured
couplings using multi-histogramming methods.  This was crucial in
obtaining a reliable estimate of the location of the peak and the peak
value.  We are indebted to Kari Rummukainen for providing us with an
appropriate code.
\bibitem{bib6}
Note that the triangular nature of the lattice we employ implies that
the phase diagram is not symmetric about the isotropic point $\kappa_1
= \kappa_2$.
\end{references}
\end{document}